\documentclass[%
 aip,
 amsmath,amssymb,
 reprint,%
]{revtex4-1}

\usepackage{graphicx}
\usepackage{dcolumn}
\usepackage{bm}

\usepackage[utf8]{inputenc}
\usepackage[T1]{fontenc}
\usepackage{mathptmx}
\usepackage{amsmath}
\usepackage{amssymb}
\usepackage{nicefrac}
\usepackage{ulem}
\usepackage{qcircuit}
\usepackage{dcolumn}   
\usepackage{mathtools}
\usepackage{braket}
\usepackage{soul}
\usepackage{cancel}
\usepackage{mathdots}
\usepackage{graphicx}
\usepackage{placeins}
\usepackage{float}
\usepackage{amsthm}
\usepackage{latexsym}
\usepackage[bb=boondox]{mathalfa}
\usepackage{etoolbox}
\usepackage{dsfont}

\makeatletter
\def\@email#1#2{%
 \endgroup
 \patchcmd{\titleblock@produce}
  {\frontmatter@RRAPformat}
  {\frontmatter@RRAPformat{\produce@RRAP{*#1\href{mailto:#2}{#2}}}\frontmatter@RRAPformat}
  {}{}
}%
\makeatother
\begin{document}

\preprint{AIP/123-QED}

\title[Low-rank decomposition for quantum simulations with complex basis functions]{Low-rank decomposition for quantum simulations with complex basis functions}
\author{Michael P. Kaicher}
\affiliation{Departamento de F\'isica Te\'orica, Universidad Complutense, 28040 Madrid, Spain}

\date{\today}

\begin{abstract}
Low-rank decompositions to reduce the Coulomb operator to a pairwise form suitable for its quantum simulation are well-known in quantum chemistry, where the underlying basis functions are real-valued. We generalize the result of Ref.~\onlinecite{motta2018low} to \textit{complex} basis functions $\psi_p(\mathbf r)\in\mathds C$ by means of the Schur decomposition and decomposing matrices into their symmetric and anti-symmetric components. This allows the application of low-rank decomposition strategies to general basis sets. 
\end{abstract}

\maketitle

\section{Introduction}
When simulating a second quantized Hamiltonian of a fermionic system with  a quantum computer, recent works employed low-rank decomposition strategies known from classical simulation algorithms to express the interaction terms as sums of squares of one-body operators \cite{whitten1973coulombic,motta2018low,chan2016matrix,peng2017highly,Matsuzawa2020,rubin2021compressing}, which allows for a simulation in terms of fermionic Gaussian unitaries and Ising-type interactions \cite{motta2018low}. With the exception of Ref.~\onlinecite{rubin2021compressing}, all such decomposition strategies rely on symmetry properties of the two-body matrix elements in the second-quantized form which result from the the underlying single-particle basis functions being real-valued. This in particular applies to molecular electronic structure type Hamiltonians, which in general do not require complex basis functions \cite{helgaker2014molecular}. 

However, other quantum systems exist that are better described by complex basis functions, where the decomposition strategy of Ref.~\onlinecite{motta2018low} can no longer be applied. One prominent example are Landau-level wave functions \cite{Landau1930} used to represent the Hamiltonian describing the fractional quantum Hall effect \cite{fano1986configuration,tsiper2002analytic,kaicher2020roadmap}. Our work shows how one can employ the low-rank representation of the Hamiltonian terms describing the two-body interaction systems represented by complex-valued single-particle basis functions. 

This paper is structured as follows. In Section~\ref{motta} we review the low-rank strategy introduced by Motta \textit{et al.}, extend their result to complex-valued basis functions in Section~\ref{low_rank}, and conclude in Section~\ref{sum}. 
\section{Low-rank representation for real-valued basis functions\label{motta}}
Second quantized time-independent Hamiltonians that describe a non-relativistic system of $N_f$ interacting fermions are typically of the following form,
\begin{align}
\hat H=\hat H_1+\hat H_2,\label{mn1}    
\end{align}
where $\hat H_1$ ($\hat H_2$) contains a linear combination of a quadratic (quartic) polynomial of fermionic annihilation and creation operators, more specifically
\begin{align}
    \hat H_1 =& \sum_{p,q=1}^{N_f}f_{p,q} \hat c_p^\dag \hat c_q,\label{mn2}\\
    \hat H_2 =& \frac{1}{2}\sum_{p,q,r,s=1}^{N_f}h_{p,q,r,s} \hat c_p^\dag \hat c_q^\dag \hat c_r\hat c_s,\label{mn3}
\end{align}
where $f$ is a two-dimensional tensor describing the one-body terms such as kinetic energy or local potentials and $h$ is a four-dimensional tensor describing the physical (e.g. Coulomb-) interaction. Here, $\hat c_p^\dag$ and $\hat c_q$ are fermionic creation and annihilation operators acting on the spin-orbitals $p$ and $q$, $f_{p,q}$ and $h_{p,q,r,s}$ are overlap integrals where the subscript $p$ coressponds to the $p$-th single particle function $\psi_p(\mathbf r)$. The factor $\frac{1}{2}$ corrects for counting the interaction between two indistinguishable fermions twice. The tensors $f$ and $h$ carry a structure that depends on the employed basis functions, the quantum system at hand, and naturally reflect the fermionic nature of the problem. 

In this section, we will present a condensed version of the method described in Ref.~\onlinecite{motta2018low}, where $\hat H$ (using a real-valued basis set) can be rewritten as the sum-of-squares of one-particle operators, which allows for a relatively simple quantum simulation of its exponential map, e.g. required to simulate its time evolution \cite{lloyd1996universal}. 

For sets of \textit{real-valued} basis functions $\psi_p(\mathbf r)\in\mathds R$, explicit expressions were given for rewriting the Hamiltonian in Eq.~\eqref{mn1} as 
\begin{align}
\hat H =& \sum_{p,q} f_{p,q} \hat c_p^\dag \hat c_q+\sum_{p,q} S_{p,q} \hat c_p^\dag \hat c_q+\frac{1}{2} \sum_{L=1}^{N_f^2}\sum_{i,j}\lambda_i^{(L)}\lambda_j^{(L)}\hat n_i^{(L)}\hat n_j^{(L)}\nonumber\\
\equiv & \hat F +  \hat S + \sum_{L}\hat V^{(L)},\label{nm4}
\end{align}
where $f$ ($S$) is a two-dimensional tensor containing all one-body contributions (from the interaction term), $\hat n_i^{(L)}$ are number operators in a rotated basis, and $\lambda_i^{(L)}$ are coefficients. In fact, the coefficients $\lambda_i^{(L)}$ are connected to the spectrum of the reshaped tensor $h$ and can be used to truncate the summation over $L$, which can lead to a significant reduction in simulation cost at low truncation error \cite{motta2018low}.  Importantly, the creation and annihilation operators in the rotated basis only satisfy the anticommutation relations for a fixed $L$ and not for $L'\neq L$. This implies that $\left[\hat V^{(L)},\hat V^{(L')}\right]\neq 0$, in general. 

As one can see from Eq.~\eqref{nm4}, the simplicity of its quantum simulation becomes evident when applying the Jordan-Wigner transformation \cite{Jordan1928}. Then, the fermionic number operators (after applying the single-particle transformation $\hat U_L$) is diagonal with no appearance of Pauli-$Z$ strings, allowing for a simple simulation by means of an Ising-type interaction.

The exponential map of the Hamiltonian operator (e.g. for performing a unitary time evolution step) can than be approximated by a Trotter step  to first order \cite{trotter1959on}, 
\begin{align}
    e^{i\Delta t\hat H} \approx& e^{i\Delta t \left(\hat F + \hat S\right)}  \prod_{L=1}^{N_f^2}\left( \hat U^{(L)} e^{i\Delta t \hat V^{(L)}}\left(\hat U^{(L)}\right)^\dag\right),\label{mn5}
\end{align}
where $\hat U^{(L)}$ are basis rotation operations which can be implemented through Givens rotations \cite{wecker2015solving}. As discussed in Ref.~\onlinecite{motta2018low}, all terms on the right-hand side can be implemented using a low-order polynomial number of Givens rotations and phase gates \footnote{The leading order depends on the underlying qubit connectivity and whether one applies approximation techniques which further reduce the gate count, see Ref.~\cite{motta2018low}.}. The purpose of this work is to give the explicit expressions for the basis rotations for the case that the underlying single-particle basis functions in Eq.~\eqref{mn1} are complex-valued. 
\section{Factorization of the interaction term for complex-valued basis functions\label{low_rank}}
In the following, we will make extensive use of flattening (also known as \textit{reshaping}) of tensors, which is why we will be rigorous with comma-separated notation in the subscripts, e.g. $h_{p,q,r,s}$ is a $(N_f\times N_f\times N_f\times N_f)$-tensor, while $h_{pq,r,s}$ denotes the reshaped $(N_f^2\times N_f\times N_f)$-tensor, where the first two dimensions are flattened.  

 We consider the interaction term 
\begin{align}
   \hat H_2 = \frac{1}{2}\sum_{p,q,r,s=1}^{N_f}h_{p,q,r,s}\hat c_p^\dag \hat c_q^\dag \hat c_r \hat c_s, \label{p1059}
\end{align}
where the two-body matrix elements are defined as 
\begin{align}
h_{p,q,r,s}=&\frac{1}{4}\left( v_{p,q,r,s} - v_{q,p,r,s} + v_{q,p,s,r} -v_{p,q,s,r} \right),\label{int0}\\
v_{p,q,r,s}=&\iint d\mathbf r_1d\mathbf r_2\psi_p^*(\mathbf r_1)\psi_q^*(\mathbf r_2)\hat V(\mathbf r_1,\mathbf r_2)\nonumber\\&\times \psi_s(\mathbf r_1)\psi_r(\mathbf r_2).\label{int2}
\end{align}
Here, $\psi_p(\mathbf r)$ describes the $p$-th basis function for a particle located at position $\mathbf r$ and $V(\mathbf r_1,\mathbf r_2)$ describes the interaction potential between particle 1 and 2, typically given by the Coulomb potential $V(\mathbf r_1,\mathbf r_2)=1/|\mathbf r_1-\mathbf r_2|$, in appropriate units.  
We will assume that the resulting tensor elements computed through the two-electron integrals are real-valued, which leads to the property $h_{p,q,r,s}=(h_{p,q,r,s})^*=h_{s,r,q,p}$. One of the more prominent examples where this holds is for the fractional quantum Hall systems mentioned in the introduction, e.g. Haldane's spherical model \cite{fano1986configuration}, or the two-dimensional disk geometry \cite{tsiper2002analytic,kaicher2020roadmap}. Note, that since we have chosen $h_{p,q,r,s}$ to possess the following symmetries, $h_{p,q,r,s}=-h_{q,p,r,s}=-h_{p,q,s,r}=h_{q,p,s,r}$, this will also translate to $h_{s,r,q,p}$, resulting in the eight-fold symmetry
\begin{align}
    &h_{p,q,r,s}=-h_{q,p,r,s}=-h_{p,q,s,r}=h_{q,p,s,r}\nonumber\\ =& h_{s,r,q,p} = -h_{r,s,q,p} = -h_{s,r,p,q} = h_{r,s,p,q}.\label{p1078}
\end{align}
We rewrite Eq.~\eqref{p1059} as 
\begin{align}
   \hat H_2 = \frac{1}{2}\sum_{p,q,r,s=1}^{N_f}h_{p,q,r,s}\hat c_p^\dag  \hat c_s \hat c_q^\dag \hat c_r + \sum_{p,r=1}^{N_f}h_{p,r}\hat c_p^\dag  \hat c_r, \label{p1060}
\end{align}
where we defined 
\begin{align}
    h_{p,r}=-\frac{1}{2}\sum_{q=1}^{N_f}h_{p,q,r,q}.\label{p1061}
\end{align}
We begin by transposing the tensor, such that indices belonging to particle 1 (here $p,s$) and particle 2 (here $q,r$) are grouped together. This is followed by flattening the tensor into a $(N_f^2 \times N_f^2)$ matrix, so that $h_{p,q,r,s}=h_{ps,qr}$. Due to the symmetry properties in Eq.~\eqref{p1078}, we know that the flattened matrix $h_{ps,qr}$ is real and symmetric, which means that we can diagonalize it by means of a Schur decomposition \cite{Schur1909}, which results in
\begin{align}
    h = O^{[h]}\Sigma^{[h]}{O^{[h]}}^T,\label{p1079}
\end{align}
where $h$ is the flattened tensor with matrix elements $h_{ps,qr}$, $O^{[h]}$ is a $(N_f^2\times N_f^2)$-real orthogonal matrix, $\Sigma^{[h]}$ is a $(N_f^2\times N_f^2)$-diagonal matrix with non-negative real-valued entries, and the superscript $[h]$ indicates that $O$ and $\Sigma$ belong here to the decomposition of $h$. With this, we can write the first term on the right-hand side of Eq.~\eqref{p1060} as 
\begin{align}
\frac{1}{2}\sum_{ps,qr}^{N_f^2}h_{ps,qr}\hat c_p^\dag  \hat c_s \hat c_q^\dag \hat c_r
=&\frac{1}{2}\sum_{ps,qr}^{N_f^2}\sum_{L=1}^{N_f^2}\left(O^{[h]}\right)_{ps,L}\left(\Sigma^{[h]}\right)_{L,L}\nonumber\\&\times\left({O^{[h]}}\right)_{qr,L}\hat c_p^\dag  \hat c_s \hat c_q^\dag \hat c_r.\label{p1080}
\end{align}
We now introduce $L$-dependent matrices $O_L^{[h]}$ whose matrix elements are given by
\begin{align}
    \left(O_L^{[h]}\right)_{p,s} = \left(O^{[h]}\right)_{ps,L},\label{p1081}
\end{align}
which can easily be obtained from \texttt{numpy.reshape()}. We can then write Eq.~\eqref{p1080} as 
\begin{align}
&\frac{1}{2}\sum_{p,q,r,s=1}^{N_f}h_{p,q,r,s}\hat c_p^\dag  \hat c_s \hat c_q^\dag \hat c_r \nonumber\\
=&\frac{1}{2}\sum_{L=1}^{N_f^2}\sum_{ps,qr}^{N_f^2}\left(\Sigma^{[h]}\right)_{L,L}\left(O^{[h]}_L\right)_{p,s}\hat c_p^\dag  \hat c_s \left({O^{[h]}_L}\right)_{q,r}\hat c_q^\dag \hat c_r,\label{p1082}
\end{align}
and Eq.~\eqref{p1079} as 
\begin{align}
    h_{ps,qr} = \sum_{L=1}^{N_f^2}\Sigma^{[h]}_{L,L}\left(O^{[h]}_L\right)_{p,s}\left(O^{[h]}_L\right)_{q,r}.\label{p1083}
\end{align}
By introducing the real-valued symmetric and anti-symmetric components of the matrix $O_L^{[h]}$,
\begin{align}
    \mathcal S^{[O_L]}=&\frac{1}{2}\left(O^{[h]}_L+\left(O^{[h]}_L\right)^T\right),\label{p1084}\\
    \mathcal A^{[O_L]}=&\frac{1}{2}\left(O^{[h]}_L-\left(O^{[h]}_L\right)^T\right),\label{p1085}
\end{align}
we can write Eq.~\eqref{p1083} as 
\begin{align}
    h_{ps,qr} =& \sum_{L=1}^{N_f^2}\Sigma^{[h]}_{L,L}\left(\mathcal S^{[O_L]}\right)_{p,s}\left(\mathcal S^{[O_L]}\right)_{q,r}\nonumber\\&+\sum_{L=1}^{N_f^2}\Sigma^{[h]}_{L,L}\left(\mathcal A^{[O_L]}\right)_{p,s}\left(\mathcal A^{[O_L]}\right)_{q,r},\label{p1086}
\end{align}
leading to an expression for the interaction term in terms of symmetric and antisymmetric matrices. One might expect two additional terms appearing in Eq.~\eqref{p1086}, namely the cross terms $\mathcal S\mathcal A$ and $\mathcal A\mathcal S$, since Eq.~\eqref{p1086} is obtained by replacing $O^{[h]}_L$ in Eq.~\eqref{p1083} with its symmetric and anti-symmetric components defined in Eqs.~\eqref{p1084}-\eqref{p1085}. However, the cross terms vanish due to the symmetry constraint $h_{ps,qr}=h_{sp,rq}$, which follows from Eq.~\eqref{p1078}.  If we denote with $L_{\mathcal{S}}$ and $L_{\mathcal{A}}$ the set of indices in $L$ which give a non-zero symmetric matrix $\mathcal S^{[O_l]}$ and $\mathcal A^{[O_l]}$, respectively, with $L_{\mathcal{S}}\cup L_{\mathcal{A}}= \{1,2,\dots,N_f^2\}$ and $L_{\mathcal{S}}\cap L_{\mathcal{A}}=\{\}$, we can write Eq.~\eqref{p1086} as 
\begin{align}
    h_{ps,qr} =& \sum_{L_{\mathcal{S}}}\Sigma^{[h]}_{L_{\mathcal{S}},L_{\mathcal{S}}}\left(\mathcal S^{[O_{L_{\mathcal{S}}}]}\right)_{p,s}\left(\mathcal S^{[O_{L_{\mathcal{S}}}]}\right)_{q,r}\nonumber\\&+\sum_{L_{\mathcal{A}}}\Sigma^{[h]}_{L_{\mathcal{A}},L_{\mathcal{A}}}\left(\mathcal A^{[O_{L_{\mathcal{A}}}]}\right)_{p,s}\left(\mathcal A^{[O_{L_{\mathcal{A}}}]}\right)_{q,r},\label{p1087}
\end{align}
which leads to 
\begin{widetext}
\begin{align}
    \frac{1}{2}\sum_{p,q,r,s=1}^{N_f}h_{p,q,r,s}\hat c_p^\dag  \hat c_s \hat c_q^\dag \hat c_r 
=&\frac{1}{2}\sum_{L_{\mathcal{S}}}\sum_{p,q,r,s=1}^{N_f}\Sigma^{[h]}_{L_{\mathcal{S}},L_{\mathcal{S}}}\left(\mathcal S^{[O_{L_{\mathcal{S}}}]}\right)_{p,s}\hat c_p^\dag  \hat c_s \left(\mathcal S^{[O_{L_{\mathcal{S}}}]}\right)_{q,r}\hat c_q^\dag \hat c_r\nonumber\\
&+\frac{1}{2}\sum_{L_{\mathcal{A}}}\sum_{p,q,r,s=1}^{N_f}\Sigma^{[h]}_{L_{\mathcal{A}},L_{\mathcal{A}}}\left(\mathcal A^{[O_{L_{\mathcal{A}}}]}\right)_{p,s}\hat c_p^\dag  \hat c_s \left(\mathcal A^{[O_{L_{\mathcal{A}}}]}\right)_{q,r}\hat c_q^\dag \hat c_r. \label{p1088}
\end{align}
\end{widetext}
We have split the Hamiltonian into the symmetric and antisymmetric components of the sliced tensor of $O^{[h]}$ that diagonalizes the Coulomb tensor in its flattened form $h_{ps,qr}$ (Eq.~\eqref{p1079}). We can now diagonalize the symmetric matrix defined in Eq.~\eqref{p1084},
\begin{align}
     \mathcal S^{[O_L]}= U^{[\mathcal S^{[O_{L}]}]}\Sigma^{[\mathcal S^{[O_{L}]}]}{U^{[\mathcal S^{[O_{L}]}]}}^\dag,\label{p1089}
\end{align}
where $U^{[\mathcal S^{[O_L]}]}$ is a unitary matrix and $\Sigma^{[\mathcal S^{[O_L]}]}$ a real-valued diagonal matrix. Similarly, one can decompose the antisymmetric matrix defined in Eq.~\eqref{p1085} into
\begin{align}
    \mathcal A^{[O_L]}= U^{[\mathcal A^{[O_{L}]}]}\Sigma^{[\mathcal A^{[O_{L}]}]}{U^{[\mathcal A^{[O_{L}]}]}}^\dag,\label{p1090}
\end{align}
where $U^{[\mathcal A^{[O_L]}]}$ is a unitary matrix and $\Sigma^{[\mathcal A^{[O_L]}]}$ is a diagonal matrix only possessing purely imaginary entries. Note, that by definition $\mathcal S^{[O_{L_{\mathcal A}}]}=\mathbb 0_{N_f}=\mathcal A^{[O_{L_{\mathcal S}}]}$.

By introducing a new set of operators
\begin{align}
    \left.{\hat b}_a^{(L)}\right.^\dag =& \begin{cases}
    \sum_{p=1}^{N_f}\left(U^{[\mathcal S^{[O_{L_{\mathcal S}}]}]}\right)_{p,a}\hat c_p^\dag,\ \text{if}\ L\in L_{\mathcal S}, \\
    \sum_{p=1}^{N_f}\left(U^{[\mathcal A^{[O_{L_{\mathcal A}}]}]}\right)_{p,a}\hat c_p^\dag,\ \text{if}\ L\in L_{\mathcal A}, 
    \end{cases}\label{p1066}\\
    \hat b_a^{(L)} =& \begin{cases}
    \sum_{p=1}^{N_f}\left({U^{[\mathcal S^{[O_{L_{\mathcal S}}]}]}}^\dag\right)_{p,a}\hat c_p,\ \text{if}\ L\in L_{\mathcal S}, \\
    \sum_{p=1}^{N_f}\left(U^{[\mathcal A^{[O_{L_{\mathcal A}}]}]}\right)_{p,a}\hat c_p,\ \text{if}\ L\in L_{\mathcal A},
    \end{cases}\label{p1067}
\end{align}
we can express the Hamiltonian in terms of $L^2$ pairwise interaction terms. Note, that the operators defined in Eqs.~\eqref{p1066}-\eqref{p1067} only fulfill the canonical anticommutation relations for a fixed $L$, but not for $L'\neq L$. By defining $\hat n_a^{(L)}=\left.\hat b_a^{(L)}\right.^{\dag}\hat b_a^{(L)}$, we arrive at an expression of the interaction terms of Eq.~\eqref{nm4} for complex-valued basis functions, 
\begin{align}
    \sum_L\hat V^{(L)} =& 
    \frac{1}{2}\sum_{L_{\mathcal{S}}}\Sigma^{[h]}_{L_{\mathcal{S}},L_{\mathcal{S}}}\sum_{a,b=1}^{N_f} \Sigma^{[\mathcal S^{[O_L]}]}_{a,a}\Sigma^{[\mathcal S^{[O_L]}]}_{b,b}\hat n_a^{(L)}\hat n_b^{(L)}\nonumber\\
    &+\frac{1}{2}\sum_{L_{\mathcal{A}}}\Sigma^{[h]}_{L_{\mathcal{A}},L_{\mathcal{A}}}\sum_{a,b=1}^{N_f} \Sigma^{[\mathcal A^{[O_L]}]}_{a,a}\Sigma^{[\mathcal A^{[O_L]}]}_{b,b}\hat n_a^{(L)}\hat n_b^{(L)}.\label{nm10}
\end{align}
Inserting the right-hand side of Eq.~\eqref{nm10} into Eq.~\eqref{nm4} gives the low-rank representation of a second quantized Hamiltonian as in Eq.~\eqref{mn1} for complex-valued basis sets. This result was derived using only the fundamental anticommutation relation and indistinguishability of fermions, and assuming that the resulting overlap integrals are real-valued. 
\section{Conclusion\label{sum}}
In this work, we presented a low-rank decomposition of a general second quantized Hamiltonian with complex-valued basis functions into a sum of squared normal operators. For complex-valued basis functions, one loses a symmetry required to employ the method of Ref.~\onlinecite{motta2018low}, and we showed how this can be overcome by expressing the transformation matrices of the Schur-decomposed reshaped tensor $h$ in terms of its symmetric and antisymmetric components, and using the fermionic nature of the indistinguishable particles to cancel cross terms that mix anti-symmetric and symmetric components. This result allows one to apply low-rank decomposition-based quantum algorithms to general basis sets. Note, that at the time of this writing, a similar result was presented in Ref.~\cite{rubin2021compressing}, which also discusses how to determine a sum-of-squares decomposition of $\hat H_2$ by a greedy search algorithm, by means of a low-depth non-orthogonal one-particle bases expansion of $\hat H_2$. Since we did not perform numerical experiments and only mentioned truncation strategies to lower the number of elements (by e.g. truncating the sums over the symmetric and anti-symmetric components $L_{\mathcal S}$ and $L_{\mathcal A}$ by introducing a truncation threshold in the eigenvalues of $\Sigma$ as in Refs.~\onlinecite{motta2018low,rubin2021compressing}), future work should focus on explicit error analysis of such strategies to our decomposition. 
\begin{acknowledgments}
MPK thanks Simon Balthasar J\"ager for discussions. MPK acknowledges funding through S2018-TCS4342 QUITEMAD-CM.
\end{acknowledgments}

\nocite{*}
\bibliography{kaicher_JMP_submission_1}

\end{document}